\documentclass[prl,twocolumn,draft,amsmath,showpacs]{revtex4}
\usepackage{graphics}\input epsf
\usepackage{times}

\begin{document}

\title{Incommensurate spin density modulation in a copper-oxide chain compound\\ with commensurate charge order}

\author{M. Raichle,$^1$ M. Reehuis,$^1$ G. Andr\'e,$^2$
L. Capogna,$^{3,4}$ M. Sofin,$^1$ M. Jansen,$^1$ and B. Keimer$^1$}

\affiliation{$^1$ Max Planck Institute for Solid State Research, Heisenbergstr. 1,
D-70569 Stuttgart, Germany}

\affiliation{$^2$ Laboratoire L\'{e}on Brillouin, CEA-CNRS,
CE-Saclay, 91191 Gif-sur-Yvette, France}

\affiliation{$^3$ CNR-INFM, CRS-SOFT and OGG Grenoble,
}

\affiliation{$^4$ Institut Laue Langevin, 6 rue J. Horowitz  BP
156 F-38042 Grenoble Cedex 9, France}

\date{\today}

\begin{abstract}
Neutron diffraction has been used to determine the magnetic
structure of Na$_8$Cu$_5$O$_{10}$, a stoichiometric compound
containing chains based on edge-sharing CuO$_4$ plaquettes. The
chains are doped with 2/5 hole per Cu site and exhibit long-range
commensurate charge order with an onset well above room temperature.
Below $T_N = 23$ K, the neutron data indicate long-range collinear
magnetic order with a spin density modulation whose propagation
vector is commensurate along and incommensurate perpendicular to the
chains. Competing interchain exchange interactions are discussed as
a possible origin of the incommensurate magnetic order.
\end{abstract}

\maketitle

The interplay between the magnetic and electric properties of copper
oxides has recently been the subject of intense research activity.
For instance, states with collinear and noncollinear magnetic order
are currently under discussion in the contexts of ferroelectricity
in undoped copper-oxide chain compounds \cite{cheong} and of the
anomalous transport properties of underdoped high temperature
superconductors \cite{luescher,fradkin}. Since theoretical methods
are well established in one dimension (1D), compounds with quasi-1D
electronic structure are particularly suitable as model systems to
obtain a detailed understanding of this interplay. However, research
on the effect of doping in copper-oxide chain compounds has been
limited by the scarcity of materials that support a significant
density of holes on the chains. Most of the attention has been
focused on the ``telephone number compounds"
(La,Sr)$_{14-x}$Ca$_x$Cu$_{24}$O$_{41}$ (LSCCO), which contain chain
and ladder systems based on edge-sharing CuO$_4$ square plaquettes
\cite{matsuda1,kataev,klingeler,schwingenschloegl,gelle,zimmermann,rusydi}.
Experiments have revealed intricate charge and spin
ordering patterns on the chain subsystem, which depend strongly on
the hole content. However, complications originating from the
presence of two distinct electronically active subsystems with
different hole concentrations and from the random potential of
substituents (for $x \neq 0$) partially mask the intrinsic behavior
of the copper-oxide chains. Moreover, recent work has shown that the
magnetic properties of this material are strongly influenced by an
incommensurate structural modulation arising from a mismatch of
different units constituting the crystal lattice
\cite{gelle,zimmermann,rusydi}. Ca$_{2+x}$Y$_{2-x}$Cu$_5$O$_{10}$
(CYCO), a class of materials containing only copper-oxide chains,
also exhibits a complicated structural modulation unrelated to
charge ordering \cite{davies,fong}. In addition, the magnetic
properties of doped chains in this material appear to be influenced
to a large extent by substitutional disorder and/or oxygen
non-stoichiometry \cite{markert}.

Na$_x$CuO$_2$, a recently synthesized family of compounds with very
low chemical disorder, offers new perspectives in this regard
\cite{sofin,horsch1,vansmaalen}. This material consists entirely of
electronically inert Na$^+$ ions and chains built of edge-sharing
CuO$_4$ plaquettes similar to those in LSCCO and CYCO (Fig. 1).
Holes donated to the chains by the Na ions form long-range ordered
superstructures that are generally incommensurate with the Na
sublattice. However, in contrast to other copper oxides with
dopable chains, incommensurate structural modulations due to purely
steric constraints are not present, so that commensurate charge
order can be established if $x$ is a rational number. By carefully
adjusting the chemical synthesis conditions, a state with $x =
1.60$, corresponding to a hole filling factor of 2/5 on the chains,
has recently been stabilized \cite{vansmaalen}. The stoichiometric
compound created in this way, Na$_8$Cu$_5$O$_{10}$, is a unique
testing ground for theories of magnetism in doped copper oxides,
without complications arising from substitutional disorder and/or
incommensurate lattice distortions.

We have determined the magnetic structure of Na$_8$Cu$_5$O$_{10}$
below its N\'eel temperature $T_N = 23$ K by neutron powder
diffraction. We find that the spins are collinear and exhibit an
incommensurate spin density modulation that is unusual for magnetic
insulators. A possible origin is a network of competing interchain
exchange interactions. An investigation of
the mechanisms stabilizing this state compared to the helicoidal
states found in the undoped analogues (Na,Li)Cu$_2$O$_2$
\cite{gippius,capogna,masuda} may provide important insights into
the competition between collinear and noncollinear magnetism in
other copper oxides of topical interest
\cite{cheong,luescher,fradkin}.

Powder samples of Na$_8$Cu$_5$O$_{10}$
with weight $\sim 4.5$ g were synthesized in a single batch as described previously \cite{sofin}.
Their magnetic susceptibility was found to be in good
agreement with prior reports \cite{sofin,horsch1}. In particular, a magnetic transition
temperature $T_N = 23$ K was obtained by analyzing the
derivative of the magnetization as a function of temperature. As
Na$_8$Cu$_5$O$_{10}$ is sensitive to air, the samples were sealed in air-tight vanadium cans, which were
loaded into a helium flow cryostat. The neutron diffraction data
were taken at the Laboratoire Leon Brillouin in Saclay, France. In
order to determine the nuclear structure, we used the
high-resolution diffractometer 3-T-2 with a neutron wavelength of
1.23 $\rm \AA$. The data for the magnetic structure determination were taken on the
high-flux cold-neutron diffractometer G-4-1 with a neutron
wavelength of 2.43 $\rm \AA$.

\begin{figure}[t]
      \begin{center}
       \leavevmode
       \epsfxsize=8.3cm
       \epsfbox{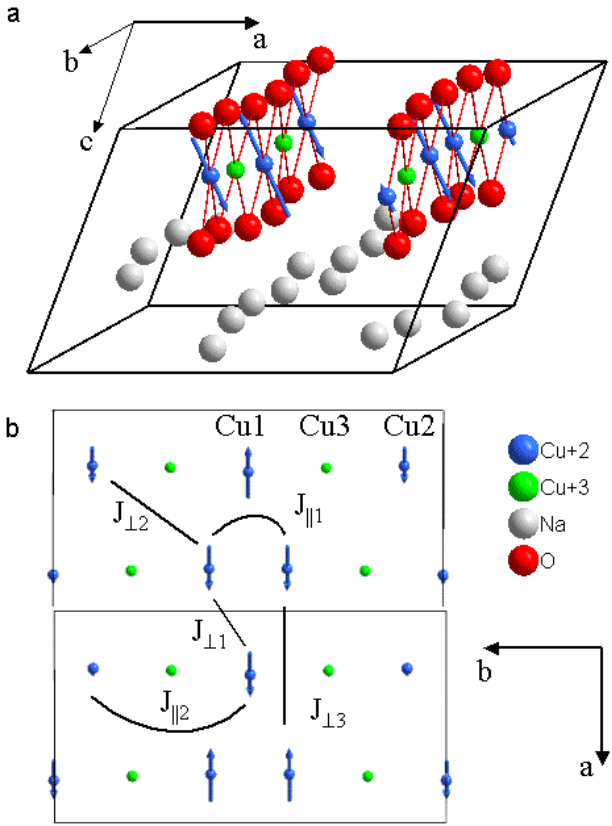}
       \caption{(a) Nuclear unit cell and spin structure of Na$_8$Cu$_5$O$_{10}$.
       (b) Cut of the lattice structure along the $ab$ plane, showing inequivalent copper sites
       and superexchange parameters, as explained in the text.}
\end{center}
\end{figure}
\begin{figure}[t]
      \begin{center}
       \leavevmode
       \epsfxsize=8.3cm
       \epsfbox{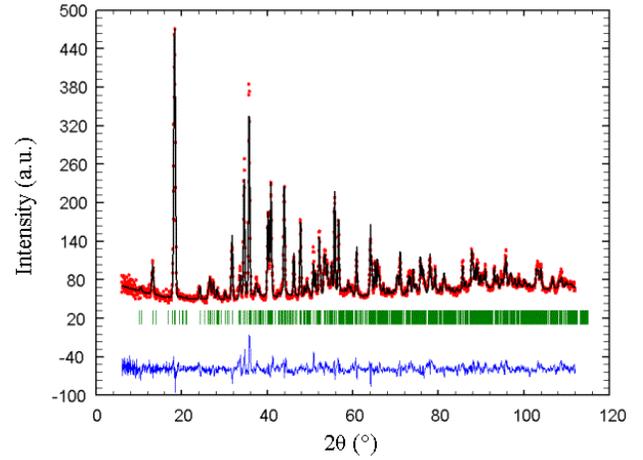}
        \caption{High-resolution neutron powder diffraction pattern of Na$_8$Cu$_5$O$_{10}$ at room temperature.
        The black line shows the result of the Rietveld refinement discussed in the text.
        The positions of nuclear Bragg reflections are indicated by green marks. The blue curve gives the difference between the calculated and measured intensities.}
\end{center}
\end{figure}

Fig. 2 shows a high-resolution powder diffraction pattern obtained
at room temperature. The nuclear intensities for both samples were
refined with the program Fullprof \cite{fullprof} on the basis of
the monoclinic space group $Cm$. The unit cell parameters $a=8.23492
\pm 0.00014 {\rm \AA}$, $b=13.92889 \pm 0.00020 {\rm \AA}$,
$c=5.71324 \pm 0.00010 {\rm \AA}$, and $\beta = 111.91 \pm 0.004
^{\circ}$ and atomic positions obtained from the refinement agree
with the results of earlier x-ray diffraction studies
\cite{sofin,vansmaalen}. The resulting diffraction pattern yields an
excellent description of the experimental data (Fig. 2), as
indicated by the goodness-of-fit-parameter $R_F = 0.0854$. This
confirms that the sample is chemically homogeneous, and that the
lattice structure is commensurate. Another sample prepared under
nominally identical conditions yielded substantially worse
refinements, and data on a sample consisting of batches of powder
material synthesized in different reactions could only be fitted by
a superposition of different charge ordering patterns. These
findings confirm that slight deviations from the ideal composition
result in incommensurately modulated structures \cite{vansmaalen}.

The atomic positions within the unit cell are displayed in Fig. 1.
The unit cell comprises ten copper ions, which are organized in
two parallel CuO$_2$ chains pointing along the $b$-axis. The
chains are separated by Na ions. Four of the copper ions in each
unit cell (Cu3 in Fig. 1b) were found to exhibit bonding patterns
characteristic of spinless Zhang-Rice singlet states
with formal valence 3+ (Ref. \onlinecite{sofin}). The Cu-O
bond lengths of the remaining six copper ions indicate a valence
state of Cu$^{2+}$ with spin 1/2. The Cu$^{2+}$ ions are
located in two inequivalent sites, which are surrounded by two
Cu$^{3+}$ ions (Cu1), and one Cu$^{3+}$ ion and one Cu$^{2+}$ ion (Cu2), respectively.
Nominally di- and trivalent copper ions are ordered in the sequence 2-2-3-2-3-2-2-3- ...
along the chains (Fig. 1). The charge order is stable up to temperatures well above room
temperature \cite{sofin,horsch1}.

Fig. 3 shows the low-angle segment of the high-flux powder pattern
measured at $T = 1.4$ K taken to determine the magnetic structure.
Two of the peaks shown vanish above  $T_N = 23$ K identified from
the anomaly in the uniform susceptibility and hence originate from
magnetic scattering. The temperature dependence of the magnetic
Bragg intensity is well described by a power-law fit without
detectable rounding near $T_N$ (Fig. 4), which indicates homogeneous
magnetic long-range order in the low-temperature phase. This
conclusion is supported by the refinement described below, which
shows that the widths of the magnetic Bragg reflections are limited
by the instrumental resolution.

The magnetic structure was refined using the software package Fullprof \cite{fullprof}
based on the magnetic structure factor


\begin{eqnarray}
\vec{F}_m(\vec{h})= \sum_{j=1} O_j f_j(\vec{h}) T_j^{iso}
\sum_s M_{js} S_{\vec{k}j} T_{js} \\ \nonumber
\exp \left\{ 2\pi i \left[ \vec{h}\{S|\vec{t}\}_s \vec{r}_j -
\Psi_{\vec{k}js} \right] \right\},
\nonumber
\end{eqnarray}

where $\vec{h}$ is the scattering vector, $\vec{k}$ is the
propagation vector of the magnetic structure, $j$ enumerates the
symmetry-inequivalent magnetic ions at positions $\vec{r}_j$ in the
magnetic unit cell, the index $s$ runs over the magnetic symmetry
operators, $M_{js}$ is an operator that transforms the Fourier
components $\vec{S}_{\vec{k}j}$ of the magnetic moments according to
a given symmetry, $\{S|\vec{t}\}_s$ generates the positions of all
symmetry-equivalent magnetic ions in the unit cell, and
$\Psi_{\vec{k}js}$ is a phase factor. The refinements are based on
the isotropic form factor $f_j(\vec{h})$ of the Cu$^{2+}$ ion. The
occupation factor $O_j$ as well as the Debye-Waller
factors, $T_j^{iso}$ and $T_{js}$, were set to unity.

\begin{figure}[t]
      \begin{center}
       \leavevmode
       \epsfxsize=8.3cm
       \epsfbox{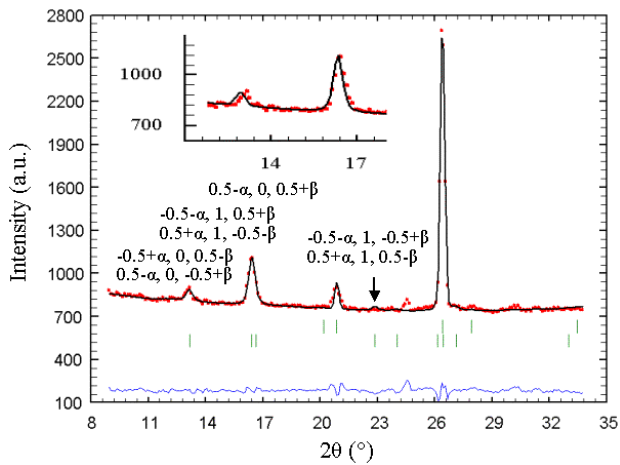}
        \caption{Low-angle segment of the high-flux powder diffraction pattern measured at 1.4K
         (red points). The black line shows the results of the Rietveld refinement discussed in the text.
         The upper (lower) green marks indicate the positions of nuclear (magnetic) Bragg reflections.
         The blue curve gives the difference between the calculated and measured intensities. The inset shows the result of a refinement based on a commensurate spin structure.}
\end{center}
\end{figure}

\begin{figure}[t]
      \begin{center}
       \leavevmode
       \epsfxsize=8cm
       \epsfbox{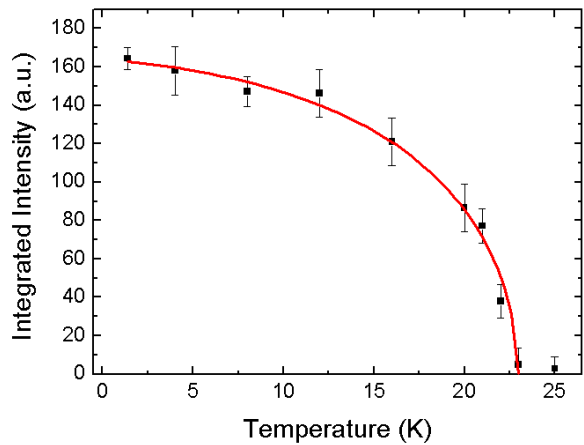}
\caption{Integrated intensities of the ($\mp 0.5 \mp \alpha$, 1,
$\pm 0.5 \pm \beta$)- and (0.5-$\alpha$, 0, 0.5+$\beta$)
magnetic Bragg reflections as a function of temperature. The line
is the result of a power-law fit.}
\end{center}
\end{figure}

The positions of the magnetic Bragg peaks indicate an approximate
doubling of the unit cell along $a$ and $c$, whereas the magnetic
and nuclear unit cells coincide along the spin-chain axis $b$. A
systematic shift away from scattering angles corresponding to
commensurate Bragg reflections (inset in Fig. 3) reveals that the
magnetic structure is incommensurate. Note that this shift can be
extracted with high confidence, because the reference lattice
parameters are determined by many Bragg reflections over a wide
range of scattering angles (Fig. 2). The propagation vector
resulting from the refinement is $(-0.5+\alpha,\, 0, \, 0.5 -
\beta)$ with $\alpha = 0.089(3)$ and $\beta = -0.030(1)$ (at 1.4 K).
The Miller indices of the magnetic Bragg reflections are shown in
Fig. 3. The asymmetric lineshape of the magnetic peak at higher
scattering angle is well explained as a consequence of the
superposition of two resolution-limited, nearly coincident
incommensurate Bragg reflections.

Although only three inequivalent magnetic Bragg reflections are
visible, the diffraction pattern imposes strong constraints on the
magnetic structure. Because of the large unit cell and the
incommensurate magnetic modulation, most possible spin arrangements
generate additional Bragg reflections with intensities well outside
the experimental error bars, where none are observed. By far the
best agreement with the data (Bragg $R=0.049$, magnetic $R=0.114$)
was obtained based on a collinear spin structure in which the two
Cu$^{2+}$ moments on Cu2 sites directly adjacent along the chains
are parallel, whereas Cu1 and Cu2 moments separated by Cu3 ions are
antiparallel. The magnetic moment on the Cu3 site was refined to
zero, consistent with the Zhang-Rice singlet state inferred from the
bond-length analysis of Ref. \onlinecite{sofin}. The incommensurate
propagation vector perpendicular to the chains modulates the
magnitude of the Cu$^{2+}$ moments. The modulation amplitude was
refined as $0.84 \pm 0.10 \mu_B$ on both Cu1 and Cu2 sites,
consistent with a spin-1/2 state. The spin direction resulting from
the refinement is $(0.86 \pm 0.39, 0, 0.92 \pm 0.07)$. An additional refinable parameter is
the phase difference $\Psi_{12} = 44.6 \pm 4.5^\circ$ of the
modulation on Cu1 and Cu2 sites.  The spin arrangement within the
nuclear unit cell is displayed in Fig. 1. The corresponding
diffraction pattern is in excellent agreement with the data (Fig.
3).

A comprehensive set of alternative collinear and noncollinear spin
structures was also tested, but the resulting refinements were
unsatisfactory. In particular, the diffraction patterns of the
circular helix structure that yields the best agreement with the
data generates prominent Bragg reflections $(\pm 0.5 \pm \alpha, 1,
\pm 0.5 \mp \beta)$ at $22.9^{\circ}$, where the Bragg intensity
vanishes within the error (arrow in Fig. 3). The corresponding
magnetic $R$-factor is 0.354, much worse than that of the collinear
state. If the refinement is generalized to include elliptical helix
structures, the length of the minor axis of the ellipse ($0.061 \pm
0.165 \mu_B$ along $b$) is consistent with zero, and the magnetic
$R$-factor does not improve significantly compared to the collinear
state. Although a small noncollinear component due to helicity or
canting cannot be ruled out, the incommensurate modulation therefore
predominantly affects the moment amplitude.

The observed spin amplitude modulation is formally analogous to spin
density waves in metallic systems, but the insulating nature and
robust charge order of Na$_8$Cu$_5$O$_{10}$ imply that it cannot
arise from a Fermi surface instability. We therefore discuss our
data in terms of superexchange interactions between local magnetic
moments, focusing first on the commensurate spin structure along the
chain axis $b$. The spin alignment along this axis indicates a
ferromagnetic nearest-neighbor ($nn$) exchange interaction $J_{\|1}$
and an antiferromagnetic next-nearest-neighbor ($nnn$) interaction
$J_{\|2}$ (Fig. 1b), in agreement with electronic structure
calculations for edge-sharing copper-oxide chains
\cite{schwingenschloegl,horsch1,mizuno,drechsler1,drechsler2,whangbo}
and with the conclusions of experiments on a variety of undoped
compounds including LiCu$_2$O$_2$ \cite{gippius,masuda} and
NaCu$_2$O$_2$ \cite{capogna}, which contain undoped chains with
similar bond lengths and angles as the ones in Na$_8$Cu$_5$O$_{10}$.
Since the Cu-O-Cu bond angle in the edge-sharing chain geometry is
close to 90$^\circ$, $J_{\|1}$ is anomalously small, and the
competing $nnn$ coupling $J_{\|2}$ is comparable or larger in
magnitude. The undoped spin systems of (Li,Na)Cu$_2$O$_2$ respond to
the resulting frustration by forming incommensurate, helical
magnetic order propagating along the chains
\cite{gippius,capogna,masuda}. In Na$_8$Cu$_5$O$_{10}$, charge
ordering lifts this frustration and gives rise to a commensurate
spin structure along the chains, as predicted by model calculations
\cite{horsch1}.

The situation is different for interactions between different
chains. For simplicity, we first ignore the small incommensurability
along $c$ and consider the magnetic bonding pattern in the
$ab$-plane (Fig. 1b), including both interactions between directly
adjacent chains within the same unit cell ($J_{\perp1}$,
$J_{\perp2}$) and interactions between $nnn$ chains ($J_{\perp3}$).
The simplest explanation for the approximate doubling of the unit
cell along $a$ is that $J_{\perp3}$ is dominant and
antiferromagnetic, leaving the interactions between $nn$ chains
frustrated.
In principle, the spin system can respond to the
frustration by establishing either noncollinear magnetic order, as
observed in (Na,Li)Cu$_2$O$_2$, or periodic spin-singlet
correlations, as found in models of frustrated and/or doped 1D
\cite{matsuda1,klingeler} and 2D \cite{fradkin} quantum
antiferromagnets. An admixture of such correlations is a possible
mechanism underlying the observed spin density modulation in
Na$_8$Cu$_5$O$_{10}$. To obtain a crude estimate of the magnitude of
the interchain interactions in the framework of this scenario, we
consider collinear classical spins coupled by sinusoidally modulated
exchange bonds with amplitudes shown in Fig. 1b. Minimization of the
exchange energy with respect to the phase shift $\Psi_{12}$ of the
modulation on the Cu1 and Cu2 sublattices then yields
$J_{\perp1}/J_{\|2}= \sin \Psi_{12}/ 2 \sin( \frac12 k_x a
+\Psi_{12}) \sim -0.7$, where $k_x$ is the component of the
incommensurate propagation vector along the $a$-axis determined by
the competing interactions between $nn$ and $nnn$ chains. While a
full quantum-mechanical calculation is required to assess the
viability of this scenario, this simple estimate indicates that the
exchange interactions between CuO$_2$ chains along $a$ are
comparable to those within the chains, as observed for other
copper-oxide chain compounds \cite{drechsler1,whangbo,boehm}. The
smaller incommensurability along $c$ suggests weaker exchange
interactions in this direction.

In view of the helicoidal states observed in (Li,Na)Cu$_2$O$_2$, the
collinear spin density modulation in Na$_8$Cu$_5$O$_{10}$ may seem
surprising. However, as other cuprates with undoped edge-sharing
chains exhibit collinear spins \cite{fong,boehm}, the energy balance
between both types of order appears to be quite subtle. This is
confirmed by ab-initio calculations \cite{whangbo,drechsler2}.
Anisotropic exchange \cite{kataev,boehm} and/or order-from-disorder
mechanisms \cite{kim} may be responsible for tipping the balance
towards collinear order in Na$_8$Cu$_5$O$_{10}$.


We thank F. Bour\'ee and B. Rieu for help during the measurements at
LLB, P. Horsch, R. K. Kremer, and P. Bourges for useful discussions,
and E. Br\"ucher for the susceptibility measurements.


\begin{thebibliography}{99}


\bibitem{cheong} See, {\it e.g.}, S. Park {\it et al.},
Phys. Rev. Lett. {\bf 98}, 057601 (2007); H.J. Xiang and M.-H.
Whangbo, {\it ibid.} {\bf 99}, 257203 (2007); Y. Naito {\it et al.},
J. Phys. Soc. Jpn. {\bf 76}, 023708 (2007).

\bibitem{luescher} See, {\it e.g.}, A. L\"uscher, A.I. Milstein, and O.P. Sushkov, Phys. Rev. Lett. {\bf 98}, 037001 (2007).

\bibitem{fradkin} See, {\it e.g.}, S. Papanikolaou, K.S. Raman, and E. Fradkin, Phys. Rev. B {\bf 75}, 094406 (2007);
M. Vojta and S. Sachdev, Phys. Rev. Lett. {\bf 83}, 3916 (1999).

\bibitem{matsuda1} M. Matsuda {\it et al.},
Phys. Rev. B {\bf 59}, 1060 (1999).

\bibitem{kataev} V. Kataev {\it et al.},
Phys. Rev. Lett. {\bf 86}, 2882 (2001).

\bibitem{klingeler} R. Klingeler {\it et al.},
Phys. Rev. B {\bf 73}, 014426 (2006).

\bibitem{schwingenschloegl} U. Schwingenschl\"ogl and C. Schuster, Phys. Rev. Lett. {\bf 99}, 237206 (2007).

\bibitem{gelle} A. Gell\'e and M.B. Lepetit, Phys. Rev. Lett. {\bf 92}, 236402 (2004).

\bibitem{zimmermann} M. v. Zimmermann {\it et al.},
Phys. Rev. B {\bf 73}, 115121 (2006).

\bibitem{rusydi}  A. Rusydi {\it et al.},
Phys. Rev. Lett. {\bf 100}, 036403 (2008).

\bibitem{davies} P.K. Davies, J. Sol. State Chem. {\bf 95}, 365 (1991).

\bibitem{fong} H. F. Fong {\it et al.},
Phys. Rev. B {\bf 59}, 6873 (1999); M. Matsuda, K. Ohyama, and M. Ohashi, J. Phys. Soc. Jpn. {\bf 68}, 269 (1999).

\bibitem{markert} M.D. Chabot and J.T. Markert, Phys. Rev. Lett. {\bf 86}, 163 (2001); M. Matsuda {\it et al.},
Phys. Rev. B {\bf 71}, 104414 (2005); K. Kudo {\it et al.}, {\it
ibid.} {\bf 71}, 104413 (2005).

\bibitem{sofin} M. Sofin {\it et al.}, J. Sol. State Chem. {\bf 178}, 3708 (2005).

\bibitem{horsch1} P. Horsch {\it et al.}, Phys. Rev. Lett. {\bf 94}, 076403 (2005).

\bibitem{vansmaalen} S. van Smaalen {\it et al.}, Acta Cryst. B {\bf 63}, 17 (2007).

\bibitem{gippius} A.A. Gippius {\it et al.}, Phys. Rev. B {\bf 70}, 020406(R) (2004).

\bibitem{capogna} L. Capogna {\it et al.},
Phys. Rev. B {\bf 71}, 140402(R) (2005).

\bibitem{masuda} T. Masuda {\it et al.}, Phys. Rev. B {\bf 72}, 014405 (2005).

\bibitem{fullprof} J. Rodriguez-Carvajal, Physica B {\bf 192}, 55 (1993).

\bibitem{mizuno} Y. Mizuno {\it et al.},
Phys. Rev. B {\bf 57}, 5326 (1998).

\bibitem{boehm} M. Boehm {\it et al.}, Europhys. Lett. {\bf 43}, 77 (1998).

\bibitem{whangbo} H.J. Xiang {\it et al.}, Phys. Rev. B {\bf 76}, 220411(R) (2007).

\bibitem{drechsler1} S.L. Drechsler {\it et al.}, Europhys. Lett. {\bf 73}, 83 (2006).

\bibitem{drechsler2} S.L. Drechsler {\it et al.}, Phys. Rev. Lett. {\bf 98}, 077202 (2007).

\bibitem{kim} Y.J. Kim {\it et al.}, Phys. Rev. Lett. {\bf 83}, 852 (1999).

%

\end{thebibliography}
\end{document}